\documentclass{article}
\usepackage{spconf,amsmath,graphicx}
\usepackage{comment}
\usepackage{hyperref}
\usepackage{multirow}
\usepackage{booktabs}
\usepackage{makecell}
\usepackage{caption}
\usepackage{xcolor}

\def\x{{\mathbf x}}

\title{Generative Pre-training for Speech with Autoregressive Predictive Coding}

\name{Yu-An Chung, James Glass}
\address{Computer Science and Artificial Intelligence Laboratory\\Massachusetts Institute of Technology\\Cambridge, MA 02139, USA\\{\small \texttt{\{andyyuan, glass\}@mit.edu}}}

\begin{document}
\ninept  % uncomment if need space

\maketitle

\begin{abstract} % 100 to 150 words
Learning meaningful and general representations from unannotated speech that are applicable to a wide range of tasks remains challenging.
In this paper we propose to use autoregressive predictive coding (APC), a recently proposed self-supervised objective, as a generative pre-training approach for learning meaningful, non-specific, and transferable speech representations.
We pre-train APC on large-scale unlabeled data and conduct transfer learning experiments on three speech applications that require different information about speech characteristics to perform well: speech recognition, speech translation, and speaker identification.
Extensive experiments show that APC not only outperforms surface features (e.g., log Mel spectrograms) and other popular representation learning methods on all three tasks, but is also effective at reducing downstream labeled data size and model parameters.
We also investigate the use of Transformers for modeling APC and find it superior to RNNs.
\end{abstract}


\section{Introduction}
The goal of speech representation learning is to find a transformation from surface features such as waveforms and spectrograms that makes high-level properties of speech (e.g., phonetic content, speaker characteristics, and even emotional cues) more accessible to downstream tasks.
Unsupervised or self-supervised objectives are especially appealing for learning representations as they can leverage unlabeled data, which are much cheaper to obtain and more scalable than datasets requiring annotation.
Representations learned via unsupervised approaches are also less likely to be biased toward a certain set of problems~\cite{chung2019unsupervised,chorowski2019unsupervised,pascual2019learning,schneider2019wav2vec,kamper2019truly}, and have more potential to be applied to a wide range of tasks.

In this paper, we aim to derive a generative pre-training approach that learns general and meaningful speech representations transferable to a variety of, potentially unknown, downstream speech tasks, where each task may require information about a different aspect of speech to perform well.
For example, phonetic content may be more crucial to speech recognition, while speaker-related applications may value the speaker information more.

Due to the required generality, we argue that it is necessary to retain in the representations as much information about the original signals as possible, and let the downstream model select which information in the representations are most useful for the task it is tackling.
However, most existing representation learning objectives~\cite{oord2018representation,milde2018unspeech,chung2018speech2vec,chen2018phonetic,hsu2017unsupervised} are designed to remove certain variabilities in speech (such as noise or speaker, depending on their design) and thus risk discarding information that could be useful for unknown downstream tasks.
Autoregressive predictive coding (APC)~\cite{chung2019unsupervised}, on the other hand, has been shown capable of learning representations that preserve information about the original signals, thus making them more {\it accessible} for downstream usage, where accessibility is defined as how linearly separable the representations are.
% define accessibility?
%\footnote{\todo{Either move this to main text or remove it} In~\cite{chung2019unsupervised} they show that the APC representations are more linear separable than the original speech signals such as log Mel spectrograms.}
This makes APC an ideal generative pre-training approach for transfer learning.
%Another advantage of APC is that its training is self-supervised, which allows it to leverage large-scale unlabeled data.
%Having this property is appealing as recent works in NLP have shown that simply increasing the amount of unlabeled pre-training data can lead to better transfer learning results~\cite{devlin2019bert,liu2019roberta}.

%\todo{related work}

The rest of the paper is organized as follows.
In Section~\ref{sec:apc} we briefly review the objective of APC and introduce two types of architectures to work as its backbone.
In Section~\ref{sec:apc_transfer} we describe how we perform transfer learning with APC.
Experiments and analysis on speech recognition, speech translation, and speaker identification are presented in Section~\ref{sec:exps}.
Finally, we conclude in Section~\ref{sec:cons} and point out some interesting future directions.

\section{Autoregressive predictive coding}
\label{sec:apc}
%We first recap the autoregressive predictive coding objective, then describe two types of architectures that work as the backbone of APC for processing sequential speech data.

\subsection{Objective}
%\todo{Illustrate APC with figure}
Autoregressive predictive coding (APC)~\cite{chung2019unsupervised} considers the sequential structures of speech and attempts to predict information about a future frame.
Inspired by the neural language modeling objective for text~\cite{mikolov2010recurrent}, which models the likelihood of a sequence of tokens to appear as a legit language, APC is trained to understand what a reasonable spectrogram should look like and encode such information in the representations.
Given a speech utterance represented as a sequence of acoustic feature vectors~(e.g., log Mel spectrograms) $\mathbf{x} = (x_{1}, x_{2}, ..., x_{N})$, APC incorporates an encoder $\mathrm{Enc}$ that encodes each frame $x_{i}$ one at a time autoregressively until the current frame $x_{k}$, and tries to predict a future frame $\x_{k + n}$ that is $n$ steps ahead of $x_{k}$.
$n\geq 1$ is meant to encourage $\mathrm{Enc}$ to infer more global structures in speech rather than exploiting local smoothness of signals.
At each time step, $\mathrm{Enc}$ produces an output prediction $y_{i}$ that has the same dimensionality as $x_{i}$.
$\mathrm{Enc}$ is optimized by minimizing the L1 loss between the predicted sequence $\mathbf{y} = (y_{1}, y_{2}, ..., y_{N})$ and the target sequence $\mathbf{t} = (t_{1}, t_{2}, ..., t_{N})$, which can be easily generated by right-shifting the input sequence $\mathbf{x}$ by $n$ time steps:
\begin{equation}
  \begin{split}
    \sum_{i = 1}^{N - n}|t_{i} - y_{i}|, t_{i} = x_{i + n}.
  \end{split}
  \label{eq:apc}
\end{equation}
Since the training target is derived from its input, APC is self-supervised and can benefit from large quantities of unlabeled data.

%We next describe two types of architectures for the encoder $\mathrm{Enc}$.

\subsection{Encoder model}
We consider two implementations of the encoder $\mathrm{Enc}$ for processing $\mathbf{x} = (x_{1}, x_{2}, ..., x_{N})$ and producing $\mathbf{y} = (y_{1}, y_{2}, ..., y_{N})$ in an autoregressive fashion: an RNN and a Transformer~\cite{vaswani2017attention}.
For the RNN, we use standard $L$-layer unidirectional GRUs~\cite{cho2014properties}:
\begin{equation}
  \begin{split}
    \mathbf{h}_{0} &= \mathbf{x},\\
    \mathbf{h}_{l} &= \mathrm{GRU}^{(l)}(\mathbf{h}_{l - 1}), \forall l\in [1, L],\\
    \mathbf{y} &= \mathbf{W}\mathbf{h}_{L},
  \end{split}
  \label{eq:rnn}
\end{equation}
where $\mathbf{W}$ projects the output of the last RNN layer $\mathbf{h}_{L}$ to the dimensionality of $\mathbf{x}$.
For an RNN-based APC, the set of trainable parameters is: $\{\mathbf{W}, \mathrm{GRU}^{(1)}, ..., \mathrm{GRU}^{(L)}\}$.

For the Transformer, similar to~\cite{liu2018generating,radford2018improving}, we consider a stack of $L$ identical {\it decoder blocks} of the original architecture~\cite{vaswani2017attention}.
Each block applies a multi-headed self-attention operation over the input sequence followed by a position-wise feedforward layer for producing the input to the next block.
We follow~\cite{vaswani2017attention} and use the sinusoidal positional encodings, which do not introduce additional parameters, to provide positional information of $\mathbf{x}$ to the model:
\begin{equation}
  \begin{split}
    \mathbf{h}_{0} &= \mathbf{W}_{\mathrm{in}}\mathbf{x} + P(\mathbf{x}),\\
    \mathbf{h}_{l} &= \mathrm{TRF}^{(l)}(\mathbf{h}_{l - 1}), \forall l\in [1, L],\\
    \mathbf{y} &= \mathbf{W}_{\mathrm{out}}\mathbf{h}_{L},
  \end{split}
  \label{eq:trf}
\end{equation}
where $\mathrm{TRF}$ stands for Transformer, $P(\cdot)$ denotes the sinusoidal encoding function, $\mathbf{W}_{\mathrm{in}}$ is an affinity that maps $\mathbf{x}$ to the dimensionality of the Transformer hidden state, and $\mathbf{W}_{\mathrm{out}}$ is another affinity that maps the final Transformer output $\mathbf{h}_{L}$ back to the dimensionality of $\mathbf{x}$.
The set of trainable parameters of a Transformer-based APC is: $\{\mathbf{W}_{\mathrm{in}}, \mathbf{W}_{\mathrm{out}}, \mathrm{TRF}^{(1)}, ..., \mathrm{TRF}^{(L)}\}$.
In practice, we tie $\mathbf{W}_{\mathrm{in}}$ and $\mathbf{W}_{\mathrm{out}}$ by setting $\mathbf{W}_{\mathrm{in}} = \mathbf{W}_{\mathrm{out}}^{T}$ as a regularization.

\section{Transfer learning with APC}
\label{sec:apc_transfer}
\subsection{Pre-training data}
We use the LibriSpeech corpus (only the speech portion)~\cite{panayotov2015librispeech} for training APC.
Specifically, the \texttt{train-clean-360} subset, which contains 360 hours of audio produced by 921 speakers in total, is used.
We use 80-dimensional log Mel spectrograms (normalized to zero mean and unit variance per speaker) as input features.
We also explore the effect of different $n$ (Equation~\ref{eq:apc}).

\subsection{Transfer learning approaches}
Once an APC feature extractor $\mathrm{Enc}$ is trained, for a downstream labeled dataset $\{(\mathbf{x}_{j}, c_{j})\}_{j = 1}^{S}$, where $(\mathbf{x}_{j}, c_{j})$ is a (feature, label) pair and $S$ denotes the training size, we transform the surface features $\{\mathbf{x}_{j}\}_{j = 1}^{S}$ (in our case, the log Mel spectrograms) into a higher-level representation with $\mathrm{Enc}$ and obtain a new dataset $\{(\mathrm{Enc}(\mathbf{x}_{j}), c_{j})\}_{j = 1}^{S}$.
Note that $c_{j}$ can be a sequence or a single value, depending on the task.

We simply take the output of the last layer of RNN or Transformer as the extracted representations, i.e., $\mathrm{Enc}(\mathbf{x}) = \mathbf{h}_{L}$ in Equations~\ref{eq:rnn} and~\ref{eq:trf}, although there are potentially better approaches that combine the internal representations across all layers~\cite{peters2018deep}.

When training a downstream model with $\{(\mathrm{Enc}(\mathbf{x}_{j}), c_{j})\}_{j = 1}^{S}$, one possibility is to keep $\mathrm{Enc}$ frozen and only optimize the model; another way is to update $\mathrm{Enc}$ as well so that the extracted representations are better adapted to the task of interest.
We examine both approaches in Section~\ref{sec:exps}.
%In Section~\ref{sec:exps} we examine both approaches, but find that the former actually works better, although the latter still outperforms the surface features.

\section{Experiments}
\label{sec:exps}
We consider three important tasks for our transfer learning experiments: (1) automatic speech recognition, (2) speaker identification, and (3) automatic speech translation.
For each task, we describe the used dataset and downstream model in their respective section.

\subsection{APC training details}
As introduced in Section~\ref{sec:apc}, we consider two architectures as the backbone of APC: RNN (Equation~\ref{eq:rnn}) and Transformer (Equation~\ref{eq:trf}), denoted as R-APC and T-APC, respectively.
For R-APC, we use 4-layer unidirectional GRUs with 512 hidden units.
Following~\cite{chung2019unsupervised}, we employ residual connections~\cite{he2016deep} between two consecutive layers.
For T-APC, we construct a 4-layer {\it decoder-only} Transformer with a hidden size of 512; each layer consists of an 8-headed self-attention module followed by a 1-layer MLP with 2048 hidden units and a GELU activation function~\cite{hendrycks2016gaussian}.
Both R-APC and T-APC are trained for 100 epochs using Adam~\cite{kingma2015adam} with a batch size of 32 and an initial learning rate of $10^{-3}$.

\subsection{Comparing methods}
We compare APC with two recently proposed self-supervised representation learning objectives: contrastive predictive coding (CPC)~\cite{oord2018representation} and problem-agnostic speech encoder (PASE)~\cite{pascual2019learning}.
%, which we briefly introduce below.

\newcommand{\myparagraph}[1]{\vspace{.4em} \noindent \textbf{#1}\ }
\myparagraph{CPC} and APC share a similar learning methodology, which is to predict information about a future frame $x_{k + n}$ based on a history $H = (x_{1}, x_{2}, ..., x_{k})$.
However, instead of trying to directly predict $x_{k + n}$ given $H$ via regression, CPC aims to learn representations containing information that are most discriminative between $x_{k + n}$ and a set of randomly sampled frames $\{\tilde{x}\}$.
The origin distribution where $\{\tilde{x}\}$ are drawn from will largely affect what information are encoded in the representations.
For example, if $\{\tilde{x}\}$ come from the same utterance as $x_{k + n}$, speaker information is likely to be discarded since they do not help distinguish $x_{k + n}$ and $\{\tilde{x}\}$.
Despite its effectiveness in tasks where the type of useful information is known (so one can select the sampling strategy accordingly), CPC might not be an ideal generative pre-training approach due to its lack of flexibility for learning general representations.

We mainly follow~\cite{oord2018representation} for implementing CPC with some modifications described in~\cite{chung2019unsupervised}.
%Readers can refer to them for more details.
As for APC, we also train CPC with the LibriSpeech \texttt{train-clean-360} subset.

\myparagraph{PASE} is a feature extractor trained by jointly optimizing multiple self-supervised objectives, where the learning target for each objective can be generated from the input signals.
Ideally, solving each task contributes to prior knowledge into the representation, resulting in a more general one that is potentially suitable for transfer learning.
% to a wide range of downstream applications.

Unfortunately, we are unable to train our own PASE using \texttt{train-clean-360}, probably due to the complexity of optimizing multiple objectives simultaneously.
Therefore, we directly use the pre-trained PASE model
%\footnote{\url{http://veu.talp.cat/models/PASE.ckpt}}
released by the authors~\cite{pascual2019learning}.
This model was trained on about 10 hours of LibriSpeech audio---according to~\cite{pascual2019learning}, they first aggregated all subsets in LibriSpeech, resulting in about 1,000 hours of audio produced by 2,484 speakers in total, then randomly selected utterances from the full set to exploit about 15 seconds of training material for each speaker.

For a fair comparison, we also train APC and CPC with approximately 10 hours of audio randomly selected from \texttt{train-clean-360}.
Note that although they are trained on about the same amount of audio, PASE has actually seen more speakers while each speaker also has fewer training material.
We add a subscript 10 to a model (e.g., CPC$_{10}$) if it is trained on only 10 hours of audio.
%\todo{If have space, add feature extraction for each model}

Below we present our transfer learning results on the three considered tasks, starting with automatic speech recognition (ASR).

\subsection{Speech recognition}
We conduct ASR experiments on the Wall Street Journal (WSJ)~\cite{paul1992design} corpus.
We follow the standard split, using 90\% of \texttt{si284} (about 72 hours) for training, the rest 10\% for development, and reporting word error rates (WER) on \texttt{dev93}.
The ASR model we use is a end-to-end, sequence-to-sequence (seq2seq) with attention architecture~\cite{chorowski2015attention} composed of an encoder and a decoder.
The encoder consists of 2 convolutional layers for downsampling the input features followed by a 4-layer bidirectional 256-dim GRU network.
The decoder is a 1-layer unidirectional 256-dim GRU network.
The seq2seq model is trained for 100 epochs using Adam with a batch size of 16 and a learning rate of $10^{-3}$.
For decoding, we use beam search with a beam size of 5.
The baseline WER using log Mel spectrograms as input features is 18.3.

\begin{table}[htbp]
  \footnotesize  % uncomment if there's enough space
  \centering
  \caption{ASR results (WER $\downarrow$) of APC with varying $n$ during pre-training and different transfer learning approaches (Frozen vs. Finetuned). log Mel is the baseline that uses log Mel spectrograms as input features. The best transfer learning result is marked in bold.}
  \begin{tabular}{lcccccc}
    \toprule
    \multirow{2}{*}{Features}  &  \multicolumn{6}{c}{$n$}\\
    \cmidrule(lr){2-7}
                     &    1   &    2   &    3   &    5   &   10   &  20\\
    \midrule
    \midrule
    log Mel          &  \multicolumn{6}{c}{18.3}\\
    \midrule
    R-APC Scratch    &  \multicolumn{6}{c}{23.2}\\
    R-APC Frozen     &  17.2  &  15.8  &  15.2  &  16.3  &  17.8  &  20.9\\
    R-APC Finetuned  &  18.2  &  17.6  &  16.9  &  18.2  &  19.7  &  21.7\\
    \midrule
    T-APC Scratch    &  \multicolumn{6}{c}{25.0}\\
    T-APC Frozen     &  19.0  &  16.1  &  14.1  &  {\bf 13.7}  &  15.4  &  21.3\\
    T-APC Finetuned  &  22.4  &  17.0  &  15.5  &  14.6  &  16.9  &  23.3\\
    \bottomrule
  \end{tabular}
  \label{tab:asr_diff_n}
\end{table}
The first experiment, presented in Table~\ref{tab:asr_diff_n}, identifies the best future time step to predict when training APC ($n$ in Equation~\ref{eq:apc}) and transfer learning approach (whether to update the pre-trained APC weights).
We also include the case where APC is randomly initialized and trained from scratch along with the seq2seq model.

From Table~\ref{tab:asr_diff_n} we observe that there exists a sweep spot when we vary $n$ for both R-APC and T-APC regardless of the transfer learning approach.
We think this is because for a small $n$, APC can exploit local smoothness in the spectrograms for predicting the target future frame (since $x_{k}$ can be very similar to $x_{k + n}$ when $n$ is small) and thus does not need to learn to encode information useful for inferring more global structures; an overly large $n$, on the other hand, makes the prediction task too challenging such that APC is unable to generalize across the training set.
The best $n$ for R-APC is 3 and for T-APC it is 5.
We also find that for all $n$, keeping pre-trained APC weights fixed (*-APC Frozen), surprisingly, works better than fine-tuning them (*-APC Finetuned), while the latter still outperforms the baseline.
Furthermore, we see that training APC from scratch along with the seq2seq model (*-APC Scratch) always performs the worst---even worse than the baseline.
With APC transfer learning, WER is reduced by more than 25\% from 18.3 to 13.7.

For the rest of the experiments we adopt R-APC Frozen with $n = 3$ and T-APC Frozen with $n = 5$.

%\todo{If have space, add nearby frame phone recognition experiment}
%O(1) access to previous hidden states, design of Transformer allows it to do so; also supported by the fact that Transformer train loss is lower

In addition to improving the performance of existing models on standard datasets, transfer learning is potentially useful for reducing the size of the downstream dataset and model needed for achieve similar performance.
The intuition is that with prior knowledge, one does not need to learn automatic feature extraction from scratch.
Being data-efficient is especially beneficial to low-resource languages with very few training pairs available, and a smaller model with competitive performance can mitigate the problem of having limited computational or storable resources.
Below we demonstrate the effectiveness of APC transfer learning in these two aspects.

\begin{table}[htbp]
  \footnotesize
  \centering
  \caption{ASR WER results with varying amounts of training data randomly sampled from \texttt{si284}. Feature extractors pre-trained with just 10 hours of LibriSpeech audio are denoted with a subscript 10.}
  \begin{tabular}{lcccccc}
    \toprule
    \multirow{2}{*}{Features}  &  \multicolumn{6}{c}{Proportion of \texttt{si284}}\\
    \cmidrule(lr){2-7}
                  &   1    &   1/2  &   1/4  &   1/8  &  1/16  &  1/32\\
    \midrule
    \midrule
    log Mel       &  18.3  &  24.1  &  33.4  &  44.6  &  66.4  &  87.7\\
    \midrule
    CPC           &  20.7  &  28.3  &  38.8  &  50.9  &  69.7  &  88.1\\
    R-APC         &  15.2  &  18.3  &  24.6  &  35.8  &  49.0  &  66.8\\
    T-APC         &  13.7  &  16.4  &  21.3  &  31.4  &  43.0  &  63.2\\
    \midrule
    PASE$_{10}$   &  20.8  &  26.6  &  32.8  &  42.1  &  58.8  &  78.6\\
    CPC$_{10}$    &  23.4  &  30.0  &  40.1  &  53.5  &  71.3  &  89.3\\
    R-APC$_{10}$  &  17.6  &  22.7  &  28.9  &  38.6  &  55.3  &  73.7\\
    T-APC$_{10}$  &  18.0  &  23.8  &  31.6  &  43.4  &  61.2  &  80.4\\
    \bottomrule
  \end{tabular}
  \label{tab:asr_data_efficiency}
\end{table}
In Table~\ref{tab:asr_data_efficiency} we compare APC with other feature extractors using varying amounts of labeled data.
For example, 1/16 means that we take only $72\times 1/16 = 4.5$ hours from \texttt{si284} for training.
We find that for all input features, there is a significant increase in WER whenever the training size is reduced by half.
When comparing R-APC and T-APC with log Mel, we see the former two always outperform the latter across all proportions, and the gap becomes larger as training size decreases.
Note that when using only half of \texttt{si284} for training, R-APC already matches the performance of log Mel trained on the full set (18.3), and T-APC even outperforms it (16.4 vs. 18.3).
In particular, we observe that T-APC always outperforms log Mel by using half of the training data log Mel uses.

When comparing the bottom half (where the feature extractors are trained on just 10 hours of audio) of Table~\ref{tab:asr_data_efficiency} with the upper part, we see that using more pre-training data is indeed helpful---performance of both CPC and APC are improved across all proportions.
This observation aligns with the findings in recent NLP literature~\cite{devlin2019bert,liu2019roberta} where having more pre-training data leads to better transfer learning results.
Finally, we see that most of the time APC outperforms CPC and PASE.
In some cases PASE is slightly better than T-APC$_{10}$ (e.g., when only 1/8 or less of \texttt{si284} is available), but is still worse than R-APC$_{10}$.

\begin{table}[htbp]
  \footnotesize
  \centering
  \caption{ASR WER results using different numbers of GRU layers for the encoder in the ASR seq2seq model.}
  \begin{tabular}{lcccc}
    \toprule
    \multirow{2}{*}{Features}  &  \multicolumn{4}{c}{Number of encoder layers}\\
    \cmidrule(lr){2-5}
                  &   1    &   2    &   3    &  4\\
    \midrule
    \midrule
    log Mel       &  28.8  &  23.5  &  20.8  &  18.3\\
    \midrule
    CPC           &  34.3  &  29.8  &  25.2  &  23.7\\
    R-APC         &  26.2  &  20.3  &  17.6  &  15.2\\
    T-APC         &  25.2  &  18.6  &  15.8  &  13.7\\
    \midrule
    PASE$_{10}$   &  29.4  &  25.7  &  22.5  &  20.8\\
    CPC$_{10}$    &  35.8  &  31.3  &  26.0  &  24.4\\
    R-APC$_{10}$  &  27.6  &  22.3  &  19.6  &  17.6\\
    T-APC$_{10}$  &  28.1  &  23.2  &  20.6  &  18.0\\
    \bottomrule
  \end{tabular}
  \label{tab:asr_model_size}
\end{table}
The next aspect we examine is to what extent can we reduce the downstream model size with transfer learning.
Specifically, in Table~\ref{tab:asr_model_size} we present the results of using different numbers of GRU layers $\in\{1, 2, 3, 4\}$ for constructing the encoder in the seq2seq model.
We see that when using the same number of layers, *-APC and *-APC$_{10}$ always outperform other features.
It is noteworthy that T-APC with just 2 layers performs similar to log Mel using 4 layers (18.6 vs. 18.3), which demonstrates the effectiveness of APC transfer learning for reducing downstream model size.

\subsection{Speech translation}
Our second task is automatic speech translation (AST), where the goal is to translate speech in one language into text in another.
We use an English-to-French translation dataset~\cite{kocabiyikoglu2018augmenting} augmented from the LibriSpeech corpus~\cite{panayotov2015librispeech} dedicated for this task.
Each data pair consists of an English waveform and its French text translation.
Following~\cite{chung2019towards}, we split the original training set, containing about 100 hours of audio, into 90\% for training and 10\% for development, and report BLEU scores~\cite{papineni2002bleu} on the dev and test sets.
The AST model we use is an RNN-based, end-to-end seq2seq with attention architecture identical to~\cite{berard2018end}.
The baseline BLEU scores using log Mel as input features on the dev and test sets are 12.5 and 12.9, respectively.

For comparison, we include the performance of the cascaded system reported in~\cite{berard2018end}.
The cascaded system pipelines an ASR module that first transcribes the input speech into text, and a machine translation (MT) module that translates the text to the target language.
A cascaded system is usually more expensive to train than an end-to-end model as it requires intermediate audio transcriptions in the source language, but serves as a strong baseline.
We also include the performance of a recently proposed Transformer-based, end-to-end AST model, dubbed S-Transformer~\cite{gangi2019adapting}, which has been shown to outperform RNN-based end-to-end model.

\begin{table}[htbp]
  \footnotesize
  \centering
  \caption{Speech translation results. BLEU scores ($\uparrow$) are reported. We also include the results of the cascaded system (ASR + MT) reported in~\cite{berard2018end} and the S-Transformer model reported in~\cite{gangi2019adapting}. Only the results on the test set are available for these two approaches.}
  \begin{tabular}{lcc}
    \toprule
    Methods        &  dev   &  test\\
    \midrule
    \midrule
    Cascaded       &   -    &  14.6\\
    S-Transformer  &   -    &  13.8\\
    \midrule
    log Mel        &  12.5  &  12.9\\
    \midrule
    CPC            &  12.1  &  12.5\\
    R-APC          &  13.5  &  13.8\\
    T-APC          &  13.7  &  14.3\\
    \midrule
    PASE$_{10}$    &  12.0  &  12.4\\
    CPC$_{10}$     &  11.8  &  12.3\\
    R-APC$_{10}$   &  13.2  &  13.7\\
    T-APC$_{10}$   &  12.8  &  13.4\\
    \bottomrule
  \end{tabular}
  \label{tab:translation}
\end{table}
From Table~\ref{tab:translation} we see that APC, regardless of how much pre-training data is used and the type of $\mathrm{Enc}$, always outperforms log Mel, CPC, and PASE on both dev and test sets.
Besides, our RNN-based model with T-APC features (14.3) outperforms S-Transformer (13.8), and is comparable with the cascaded system (14.6).

\subsection{Speaker identification}
\begin{table}[htbp]
  \footnotesize
  \centering
  \caption{Speaker ID results. Accuracies ($\uparrow$) are reported.}
  \begin{tabular}{lcccccc}
    \toprule
    \multirow{2}{*}{Features}  &  \multicolumn{6}{c}{Number of utterances per speaker seen in training}\\
    \cmidrule(lr){2-7}
                  &   1    &   5    &   10   &   20   &   50   &  full (130 in avg.)\\
    \midrule
    \midrule
    log Mel       &  8.7  &  43.7  &  60.4  &  70.5  &  87.4  &  96.1\\
    \midrule
    CPC           &  13.0  &  45.5  &  65.8  &  75.9  &  89.3  &  96.5\\
    R-APC         &  17.2  &  56.9  &  73.3  &  87.4  &  95.1  &  99.0\\
    T-APC         &  17.6  &  58.6  &  74.4  &  87.8  &  96.3  &  99.1\\
    \midrule
    PASE$_{10}$   &  12.5  &  48.6  &  64.8  &  79.6  &  92.6  &  96.7\\
    CPC$_{10}$    &  11.7  &  44.9  &  63.2  &  74.6  &  88.3  &  95.8\\
    R-APC$_{10}$  &  14.3  &  54.4  &  72.3  &  87.1  &  95.0  &  98.9\\
    T-APC$_{10}$  &  13.5  &  49.2  &  70.5  &  82.8  &  92.4  &  98.0\\
    \bottomrule
  \end{tabular}
  \label{tab:spk_id}
\end{table}
Our final task, speaker identification (SID), examines how much transferable speaker information is captured by the representations learned by different objectives.
We use WSJ for our SID experiments.
We split \texttt{si284} into 80\% for training, 10\% for development, and 10\% for testing.
The task is equivalent to a 259-speaker classification problem.
Features are fed into a 1-layer GRU network with a Softmax layer appended on top of the output of the last time step, which is optimized by minimizing the negative log-likelihood across the training set.
We investigate settings where different amounts of utterances per speaker are used for training---in the most extreme case only one utterance per speaker is available.
Exploring such one- or few-shot learning scenarios is especially interesting as it is closer to the real world where, for instance, a speech application on a personal device needs to quickly adapt to user-specific features with just a few input samples for better user experience.

From Table~\ref{tab:spk_id} we see that APC representations contain more transferable speaker information than all the other features, almost always outperforming them regardless of how many utterances per speaker are seen during training.
It is noteworthy that T-APC is almost twice as good as log Mel (17.6 vs. 8.7) in one-shot learning.

\section{Conclusions}
\label{sec:cons}
% Conclusions
We demonstrate that autoregressive predictive coding (APC) is an effective generative pre-training objective for transfer learning to a wide range of speech tasks.
We use a Transformer to model APC and empirically show that it is more effective than an RNN used in~\cite{chung2019unsupervised}.
On speech recognition (ASR), speech translation, and speaker identification (SID), representations learned by APC consistently and, mostly, significantly outperform log Mel spectrograms and representations learned by other objectives such as CPC~\cite{oord2018representation} and PASE~\cite{pascual2019learning}.
We also investigate the data efficiency and model efficiency aspects on ASR and SID, and show that APC representations are the most effective among all comparing methods at reducing downstream labeled data size and model parameters.

% IF YOU SAY "SIGNIFICANT" YOU NEED TO QUANTIFY IT

% Future work
There are many interesting directions for future work.
In our experiments, we find that keeping APC weights frozen works better than updating them when training on downstream dataset.
However, we believe that the latter is more ideal for transfer learning as it adapts the extracted representations toward the target task.
More sophisticated techniques for fine-tuning~\cite{howard2018universal} could be used.
Regarding the backbone architecture of APC, the Transformer model can be potentially improved by modifying the way we inject positional information~\cite{mohamed2019transformers,irie2019language}.
Additionally, as hinted by recent NLP research~\cite{devlin2019bert,liu2019roberta}, training APC on more unlabeled data is also a promising way for improving the transfer learning results.
Finally, we are interested in exploring the usage of APC in other speech applications such as speech synthesis, where pre-training and transfer learning have already achieved some success~\cite{chung2019semi,chen2019end,hayashi2019pre,fang2019towards,jia2018transfer}.

%\newpage  % for testing reference page

\bibliographystyle{IEEEbib}
\bibliography{strings,refs}

\end{document}